\documentclass{article}

\def\abstract#1{\vskip 7mm 
        \begin{center}{\large Abstract}\par \smallskip
                \begin{minipage}[c]{12cm}
                        \small #1
                \end{minipage}
        \end{center}
}
\def\title#1{\begin{center}{\Large\bf #1}\end{center}}
\def\author#1{\vskip 5mm \begin{center}{#1}\end{center}}
\def\address#1{\begin{center}{\it #1}\end{center}}

\usepackage{latexsym}
\usepackage{amssymb}
\usepackage{amsthm}
\usepackage{amsmath}
\usepackage{amsfonts}
\usepackage[dvips]{graphicx}

\newcommand{\sikib}{\begin{eqnarray}}
\newcommand{\sikie}{\end{eqnarray}}
\newcommand{\sikibnon}{\begin{eqnarray*}}
\newcommand{\sikienon}{\end{eqnarray*}}
\newcommand{\cenb}{\begin{center}}
\newcommand{\cene}{\end{center}}

\newcommand{\gin}[1]{g_{in}(\vec{#1})}
\newcommand{\gout}[1]{g_{out}(\vec{#1})}

\makeatletter
\@ifundefined{lesssim}{}{}
\@ifundefined{gtrsim}{}{}
\def\vereq#1#2{\lower3pt\vbox{\baselineskip1.5pt \lineskip1.5pt
\ialign{$\m@th#1\hfill##\hfil$\crcr#2\crcr\sim\crcr}}}
\makeatother

\begin{document}

\title{%
Black Hole Evaporation and Nonequilibrium Thermodynamics for a Radiation Field
}
\author{%
  Hiromi Saida\footnote{E-mail:saida@daido-it.ac.jp},
}
\address{%
 Department of Physics, Daido Institute of Technology, Nagoya 457-8530, Japan
}

\abstract{
When a black hole is put in an {\it empty} space (zero temperature space) on which there is no matter except the matter of the Hawking radiation (Hawking field), then an outgoing energy flow from the black hole into the empty space exists. By the way, an equilibrium between two arbitrary systems can not allow the existence of an energy (heat) flow from one system to another. Consequently, in the case of a black hole evaporation in the empty space, the Hawking field should be in a nonequilibrium state. Hence the total behaviour of the evaporation, for example the time evolution of the total entropy, should be analysed with a nonequilibrium thermodynamics for the Hawking field. This manuscript explains briefly the way of constructing a nonequilibrium thermodynamic theory for a radiation field, and apply it to a simplified model of a black hole evaporation to calculate the time evolution of the total entropy.
}

\section{Introduction}
\label{sec-intro}

The black hole evaporation due to the Hawking radiation is one of the interesting dynamical behaviours of a black hole. However, in comparison with the study on the gravitational collapse (formation of a black hole), fewer details of the black hole evaporation has been revealed so far. This manuscript aims to analyse some detail property of the black hole evaporation in four dimensions.
\footnote{The two dimensional dilatonic theory has provided an exact solution for a black hole evaporation. However in this manuscript, we are interested in the four dimensional black hole evaporation in the framework of the Einstein gravity.} 
Hereafter we set the fundamental constants unity, $\hbar = G = c = k_B = 1$. 

On the study of the black hole evaporation, the so-called generalised second law (GSL) of the black hole thermodynamics has been one of the interesting issues, and it has become a general agreement today that a black hole is a thermodynamic object whose equations of states are
\sikib
 E_g = \frac{1}{8 \pi T_g} = \frac{R_g}{2} \quad , \quad
 S_g = \frac{1}{16 \pi T_g^{\,2}} = \pi R_g^{\,2} \, ,
\label{eq-eos}
\sikie
where the Schwarzschild black hole is considered for simplicity, $R_g$ is the radius, $E_g$ the energy, $T_g$ the temperature and $S_g$ is the entropy of the black hole. The GSL describes, for example, that the total entropy $S_{tot}$ of the total system which consists of the black hole and the matter field of the Hawking radiation is given by a simple sum $S_{tot} = S_g + S_m$, where $S_m$ is the entropy of the matter field, then $S_{tot}$ increases monotonously, $dS_{tot} > 0$, during the black hole evaporation process. 

When a black hole is put in an {\it empty} space (zero temperature space) on which there is no matter field except that of the Hawking radiation, then an outgoing energy flux $J$ from the black hole into the empty space exists during the evaporation process. Since it seems appropriate during the evaporation process to assume that, from the thermodynamic viewpoint, the black hole itself passes a sequence of equilibrium states characterised by the equation (\ref{eq-eos}), then the flux $J$ is given by the Stefan-Boltzmann law,
\sikib
 J = \sigma \, T_g^{\,4} \, A_g \, ,
\label{eq-flux}
\sikie
where $\sigma$ is the Stefan-Boltzmann constant and $A_g = 4 \pi R_g^{\,2}$ is the area of the black hole. By the way, an equilibrium between two arbitrary systems can not allow the existence of an energy (heat) flow from one system to another. Consequently, in the case of a black hole evaporation in the empty space, while the black hole passes a sequence of equilibrium states, the matter field of the Hawking radiation should pass a sequence of nonequilibrium states, where each nonequilibrium state in the sequence of the field's states should have the energy flux $J$. That is, the total behaviour of the evaporation, for example the time evolution of the total entropy $S_{tot}$, should be analysed with a nonequilibrium thermodynamics for the matter field of the Hawking radiation. However, because a definition of a nonequilibrium entropy has not generally been constructed in a consistent way, most of the works concerning the GSL treat an equilibrium between a black hole and a heat bath surrounding it, or assume the existence of a nonequilibrium entropy of matter fields. 

Our main point in this manuscript is to deal with a concrete form of a nonequilibrium entropy carried and produced by the Hawking radiation. For simplicity, we consider the photons as the representative of particles of the Hawking radiation. That is, a massless free field of two helicities are used as the matter field of the Hawking radiation. 

Further, in order to pick up the nonequilibrium effects of the matter field (photon field), we consider a simplified toy model: put a spherical {\it black body} satisfying the equation (\ref{eq-eos}) in a Minkowski spacetime, then consider the relaxation process of the black body due to the emission of the thermal radiation. This relaxation process picks up the nonequilibrium effects of the Hawking radiation, but neglects the general relativistic effects like the lens effects on photons emitted at the horizon in off-radial directions.

\section{Steady state thermodynamics for a radiation field}

\begin{figure}[t]
 \begin{center}
 \includegraphics[height=25mm]{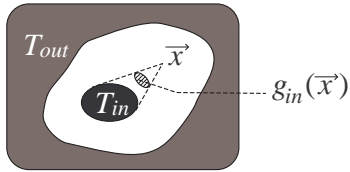}
 \end{center}
\caption{Pick up only the nonequilibrium effects of the Hawking radiation.}
\label{pic}
\end{figure}

As mentioned above, we want to consider a simplified model that a black body is put in an infinitely large Minkowski space. However, as a middle step before proceeding to such simple model, we consider a more simplified model shown in figure \ref{pic}: make a vacuum cavity in a large black body of temperature $T_{out}$ and put another (smaller) black body of temperature $T_{in} \, ( \neq T_{out} )$. Hereafter we set $T_{in} > T_{out}$ without loss of generality. Then, there should exist an energy flow from the inner black body to the outer black body due to the thermal radiation emitted by them, and the relaxation process leads the whole system to a total equilibrium where the two black bodies and a radiation field between them have the same temperature. 

Consider the case satisfying the following two conditions: (C1) each body passes a sequence of equilibrium states during the relaxation process, and (C2) the volume of the cavity is so small that the speed of light is approximated as infinity during the relaxation process. Then the radiation field sandwiched by the two black bodies should pass a sequence of nonequilibrium states. Each nonequilibrium state in the sequence has the energy flux $J_{rad} = \sigma \, (\, T_{in} - T_{out} \, ) \, A_{in}$, where $A_{in}$ is the surface area of the inner black body. This denotes that it is enough for us to consider a set (state space) of the {\it steady states} of the radiation field whose nonequilibrium nature arises by the steady energy flow $J_{rad}$.
\footnote{The terminology "steady state" means the macroscopically stationary nonequilibrium state.} 
That is, the sequence of the nonequilibrium states of the radiation field during the relaxation process lies in the set (state space) of the steady states. Hence we search for a consistent definition of the steady state entropy of the radiation field. When such an entropy of the radiation field is obtained, the entropy during the relaxation process of the two black bodies and the radiation field can be explicitly expressed. Further, by extending the definition of the steady state entropy into the case of an infinitely large cavity with $T_{out} = 0$, it is expected that the total entropy $S_{tot}$ of the simplified model of the black hole evaporation discussed in previous section can be analysed explicitly. 

The author H.S. has already constructed a consistent thermodynamic framework of the steady states for a radiation field \cite{ref-sst}. The outline of the construction of the entropy in the steady state thermodynamics for a radiation field is as follows:
\begin{enumerate}
\item
 Refer the Landau-Lifshitz type definition of a nonequilibrium entropy for a bosonic gas \cite{ref-ll}
\sikib
 S_{boson} =
  \int \frac{dp^3}{(2\pi)^3} dx^3 \, g_{\vec{p},\vec{x}}
  \left[\, \left(\, 1 + N_{\vec{p},\vec{x}} \,\right) \,
             \ln\left(\, 1 + N_{\vec{p},\vec{x}} \,\right)
         - N_{\vec{p},\vec{x}} \, \ln N_{\vec{p},\vec{x}}
  \,\right] \, ,
\label{eq-ll}
\sikie
where $\vec{p}$ is the momentum of a photon, $\vec{x}$ is a spatial point, and $g_{\vec{p},\vec{x}}$ and $N_{\vec{p},\vec{x}}$ are respectively the number of states and the average number of (bosonic) particles at a point $(\vec{p},\vec{x})$ in the phase space of the boson gas.
\footnote{It has also been shown in reference \cite{ref-ll} that the maximisation of $S_{boson}$ for an isolated system ($\delta S_{boson} = 0$) gives the equilibrium Bose distribution. This is frequently refereed in many works on nonequilibrium systems as the "H-theorem". But in reference \cite{ref-ll}, the concrete forms of $g_{\vec{p},\vec{x}}$ and $N_{\vec{p},\vec{x}}$ are not specified, since an arbitrary system is considered. }
\item
 In our system, the radiation field is sandwiched by two perfect black body. Then we can determine $g_{\vec{p},\vec{x}}$ and $N_{\vec{p},\vec{x}}$ as
\sikibnon
 g_{\vec{p},\vec{x}} = 2 \quad , \quad
 N_{\vec{p},\vec{x}} = \frac{1}{\exp[\omega / T(\vec{p},\vec{x})] - 1}
\sikienon
where the frequency of a photon $\omega = p$, and $T(\vec{p},\vec{x})$ is given by
\sikibnon
 T(\vec{p},\vec{x}) =
 \begin{cases}
  T_{in}  & \text{for} \,\,\, \vec{p} = \vec{p}_{in} \,\,\, \text{at $\vec{x}$} \\
  T_{out} & \text{for} \,\,\, \vec{p} = \vec{p}_{out} \,\,\, \text{at $\vec{x}$}
 \end{cases} \, ,
\sikienon
where $\vec{p}_{in}$ is the photon emitted by the inner black body and $\vec{p}_{out}$ by the outer black body. The directions from which the photons of $\vec{p}_{in}$ and $\vec{p}_{out}$ can come to a point $\vec{x}$ vary from point to point. Therefor the $\vec{x}$-dependence of $T(\vec{p},\vec{x})$ arises.
\item
 From above two steps, we obtain the steady state entropy for a radiation field, $S_{rad}$,
\sikib
 S_{rad} = \int dx^3 s_{rad}(\vec{x}) \quad , \quad
 s_{rad}(\vec{x}) = \frac{16 \, \sigma}{3} \,
  \left(\, \gin{x}\, T_{in}^{\,3} + \gout{x}\, T_{out}^{\,3} \,\right) \, ,
\label{eq-sst.entropy}
\sikie
where $\gin{x}$ is the solid angle (divided by $4 \pi$) covered by the direction of $\vec{p}_{in}$ at $\vec{x}$ as shown in figure \ref{pic}, and $\gout{x}$ is similarly defined with $\vec{p}_{out}$.
\item
 Further, with defining the steady state internal energy $E_{rad} = 2 \int dp^3 dx^3 \, \omega \, N_{\vec{p},\vec{x}}$ and the other variables like the free energy with somewhat careful discussions, it has been already checked that the 0th, 1st, 2nd and 3rd laws of the ordinary equilibrium thermodynamics are extended to include the steady states of a radiation field. That is, the steady state thermodynamics (SST) for a radiation field has already been constructed in a consistent way.
\end{enumerate}

\section{Time evolution of $S_{tot}$}

We extend the SST entropy (\ref{eq-sst.entropy}) to the case that the volume of the cavity is infinity, the outer temperature is zero $T_{out} = 0$ and the equations of states for the inner black body are given by (\ref{eq-eos}). As mentioned in the section \ref{sec-intro}, this case corresponds to the simplified model of the black hole evaporation in an "empty" space with neglecting the gravitational effect like the lens effects on photons emitted in off-radial directions. Further the inner black body is the representative of the black hole, so we set $T_{in} = T_g$. The time evolution of $T_g$ is given with equations (\ref{eq-eos}) and (\ref{eq-flux}),
\sikib
 \frac{d E_g}{dt} = - J \quad \Rightarrow \quad
 T_g(t) = T_0 \, \left( 1 - 6 \sigma \, T_0^{\,3} \, t \right)^{-1/3}
        = \frac{1}{4 \pi \, R_g(t)} \, ,
\label{eq-temperature}
\sikie
where $T_0 = T_g(0)$ is the initial condition. 

The other points we should consider are the effects of the retarded time of photons travelling in an infinitely large space, and the Doppler effects due to the "shrinkage" speed of the surface of the inner black body (the surface of the black hole). With taking these effects into acount, the total entropy is obtained to be
\sikib
 S_{tot}(t) = S_g(t) + 4 \pi \, \int_{R_g(t)}^{r_p(t)} dr\,r^2 \, s_{rad}(r,t) \, ,
\label{eq-total.entropy}
\sikie
where $r$ is the areal radius, $r_p(t)$ is the radius of the "wave front" of emitted photons
\footnote{Here we assume that the evaporation starts at $t=0$. Then the photons emitted at $t=0$ forms the boundary of the region where the radiation field exists. We call this boundary the "wave front".} 
and the time dependence of $S_g(t)$ and $s_{rad}(r,t)$ arise from (\ref{eq-temperature}).

\section{Conclusion and next step}

\begin{figure}[t]
 \begin{center}
 \includegraphics[height=85mm]{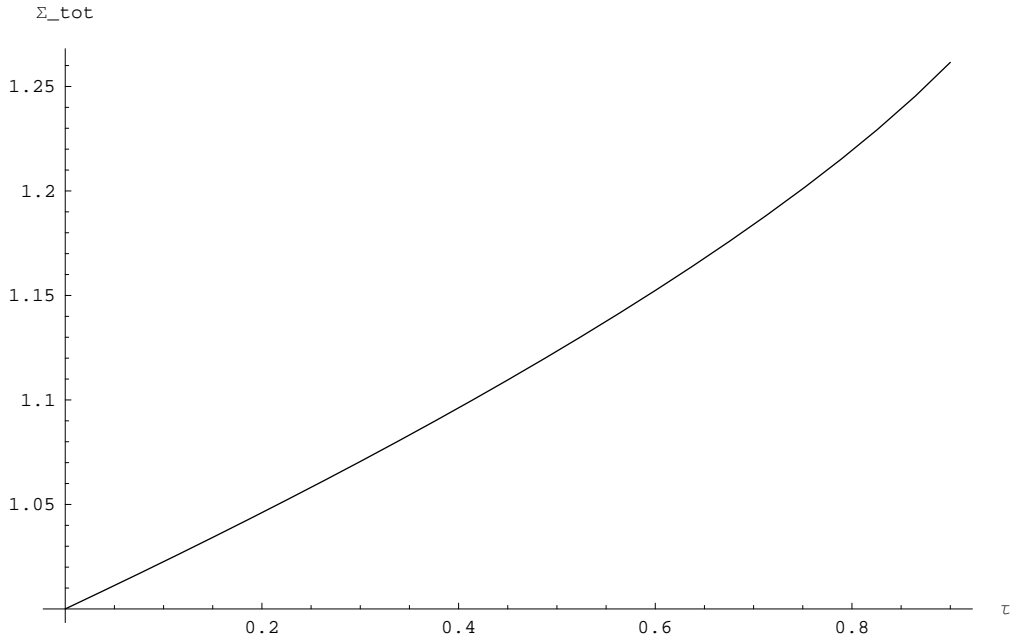}
 \end{center}
\caption{Time evolution of $\Sigma_{tot} = S_{tot}(t)/S_{tot}(0)$.}
\label{pic-2}
\end{figure}

Figure \ref{pic-2} are the plot of $\Sigma_{tot} \equiv S_{tot}(t)/S_{tot}(0)$ given by (\ref{eq-total.entropy}) with setting $6 \sigma \, T_0^{\,3} = 1$. As far as the author knows, this is the first example to show a time evolution of the total entropy $S_{tot}$. And as is expected, this result supports the validity of the GSL. For the next step along the strategy discussed in this manuscript, it is necessary to extend our result to include the gravitational effects like the lens effects on photons emitted in off-radial directions at the horizon. Further, since we has assumed that the Stefan-Boltzmann law (\ref{eq-flux}) holds instead of solving the Einstein equation for the evaporating black hole, it should also be searched how the background spacetime of the black hole evaporation is described.

\end{document}